\documentclass[12pt]{article}

\usepackage{amsmath,amssymb}
\usepackage{amssymb}

\DeclareMathOperator{\Tr}{Tr}
\DeclareMathOperator{\tr}{tr}
\DeclareMathOperator{\Det}{Det}

\newcommand{\EA}{\text{EA}}
\newcommand{\Ee}{\text{E}}
\newcommand{\Si}{\text{Si}}
\newcommand{\KK}{\text{K}}

\begin{document}

\begin{titlepage}

\vspace*{1.4cm}

\title{Regularization of Propagators and Logarithms in Background
    Field Method in 4-dimensions}
\begin{center}
    \Large{Regularization of Propagators and Logarithms \\
           in the Background Field Method in 4-dimensions}
\end{center}
\vspace{1.2cm}
\begin{center}
    {\large\bf T.~A.~Bolokhov}

\vspace{0.6cm}
    {\it St.Petersburg Department of V.\,A.\,Steklov Mathematical Institute\\ 
         Russian Academy of Sciences\\
	 27 Fontanka, St.Petersburg, Russia 191023}
\end{center}

\vspace{0.8cm}



\centerline{\large\bf Abstract}

\vspace{2mm}

    The determinant and higher loop terms,
    usually treated with the Pauli-Villars and higher covariant derivatives methods,
    in the background field method
    can hardly be regularized simultaneously.
    At the same time we observe that introduction of a scalar multiplier
    in front of the quadratic form, which is equivalent to a change of the
    measure in the functional integral, influences only the determinant part of
    the effective action.
    This allows one to choose the integration measure and the function in the
    regularized propagator in such a way as to make all terms in the
    expansion finite.

\end{titlepage}

\newpage
    Originally introduced in 
\cite{DeWitt, Abbott},
    the background field method significantly simplifies calculation of the effective action and the
$ \beta $-function
    in quantum field models.
    In a general case this method implies taking a functional integral over
    quantum fluctuations
$ b $
    around a background field
$ B $:
\begin{equation*}
    Z(B) = \int \exp \{i S(B,b)\} \prod \delta b ,
\end{equation*}
    where
$ S(B,b) $
    is the modified action of the classical theory.
    Normally, this action is constructed from the classical one by substituting
$ B+b $
    as its argument.
    If, however, the theory contains an additional symmetry then 
    gauge-fixing terms should be added
\begin{equation*}
    S(B,b) = S_{\text{cl}}(B+b) + S_{\text{gauge}}(B,b)\,,
\end{equation*}
    and in this case $ S $ will not be just
    a function of the sum.
    Let us assume that the expansion of the classical action
    around the zero of its argument
    consists of a finite number of terms.
    Then, after introducing a coupling constant
$ g $
    and replacing
\begin{equation*}
    b \to g b , \quad S \to \frac{1}{g^{2}} S\,,
\end{equation*}
    the modified action reads as
\begin{align}
\nonumber
    \frac{1}{g^{2}}S(B,gb) = & \frac{1}{g^{2}} S_{\text{cl}}(B)
	+ \frac{1}{g}V_{1}b + \frac{1}{2} bMb + gV_{3}b^{3} + \ldots
	    + g^{N-2}V_{N}b^{N} = \\
\label{Sexp}
    = & \frac{1}{g^{2}} S_{\text{cl}}(B)
	+ \frac{1}{g}V_{1}b + \frac{1}{2} bMb + g S_{\text{Int}} .
\end{align}
    Here and further on we assume that fields and vertices (the interaction
    points
$ V $)
    may carry both vector indices and the indices related to the internal symmetry, and
    also that the integration variable
$ b $
    incorporates the auxiliary (ghost) fields.

    Instead of calculating
$ Z(B) $,
    it is more useful to calculate its normalized logarithm
    which is called the effective action.
    By taking constant
$ g $
    small, the effective action can be represented
    as a sum of connected Feynman diagrams, in which the propagators
$ M^{-1} $
    and the vertices
$ V_{k} $
    now depend on the background field
$ B $:
\begin{align}
\nonumber
    \EA&(B) = \ln Z(B) - \ln Z(0) = \\
\label{EAint}
	= &\ln \int \exp \bigl\{ \frac{i}{g^{2}} S_{\text{cl}}(B)
	+ \frac{i}{g}V_{1}b + \frac{i}{2} bMb + i g S_{\text{Int}}
	\bigr\} \prod \delta b - \ln Z(0)= \\
\label{EAexp}
    = &\frac{i}{g^{2}} S_{\text{cl}} 
	+ \frac{i}{2} \Tr (\ln M^{-1}(B) - \ln M^{-1}(0))
    + ig^{2} (\text{2 Loops}) + \ldots .
\end{align}
    Here we have eliminated the contribution of the linear term
$ \frac{1}{g}V_{1}b $,
    which would have generated an infinite series of additional terms
    in each step of the expansion in $ g^2 $.
    This is justified, if we 
    impose a constraint on the field
$ B $
    called the quantum equation of motion 
    (to first approximation it coincides with the classical
    equation of motion)
    that eliminates the contribution of one-particle reducible
    diagrams
\cite{FSep}.

    Now the sum
\eqref{EAexp}
    contains divergent integrals. 
    In particular, the trace of the logarithm is divergent,
    while the loop expansion produces
    multiple divergent integrals of the type
\begin{equation}
\label{divint}
    \int \bigl(M^{-1}(x,y)\bigr)^{2} d^{4}(x-y) \simeq \frac{1}{(4\pi^{2})^{2}}
	\int \frac{d^{4}(x-y)}{(x-y)^{4}}
\end{equation}
    as well as others (here and further we say a loop diagram whenever the 
    diagram has more than one loops).
    The goal of the regularization procedure is to change the expression in
    the functional integral
\eqref{EAint}
    in such a way that all terms in the sum
\eqref{EAexp}
    become finite.
    It is of course necessary to insist that the integral
\eqref{EAint}
    restores its initial form when the parameter that describes this
    change is taken to a certain value.
    Then, by considering the limits of the effective action upon 
    different behaviour of the coupling constant and regularization parameter, 
    one can set up the problem of renormalization.

    The above approach to the background field method is described in
\cite{LFC}.
    Its advantage is that it allows for a direct control over the symmetry
    of the theory via the dependence of the coefficients
$ V_{k} $ and
$ M $
    in the integral
\eqref{EAint}
    on the background field
$ B $.
    There is practically but one regularization scheme compatible
    with the above prescription at two loops and beyond
    --- dimensional regularization
\cite{JO}.
    In the latter approach the action
$ S $
    is transferred into a space of dimension
$ 4-\epsilon $,
    where dimensionless
$ \epsilon $
    acts as a regularization parameter.
    Then, the trace of the logarithm of the propagator and the divergent integrals
    of the type \eqref{divint}
    turn into expansions in inverse powers of
$ \epsilon $
    ({\it i.e.} into Laurent series).

    In this paper we discuss a natural question of whether it is possible
    to, instead, regularize the integral in Eq.~\eqref{EAint}
    in the original 4-dimensional Euclidean space, 
    by changing the propagator
$ M^{-1} $ 
    (which is obtained from the operator $ M $ of the quadratic form) 
    into an appropriately chosen function of
$ M $:
\begin{align*}
    M^{-1} & \to  r(M,\Lambda) , \quad r(M,\Lambda)
	\stackrel{\Lambda\to\infty}{\longrightarrow} M^{-1} , \\
    M & \to  r^{-1}(M, \Lambda) , \\
    \ln M^{-1} &\to \ln r(M,\Lambda) ,
\end{align*}
    where
$ \Lambda $
    is the regularization parameter.
    Using the Yang-Mills field as an example we argue
    that the loop divergences and the logarithm trace divergence
    are in fact inter-related and cannot be regularized by a single
    function 
$ r $
    with an analytic behaviour, at least not by one from within the class of Laplace
    transformations.
    However, as we demonstrate in Section 2 this can still be done by
    a step-like function. 
    This approach --- the restriction of the integration domain ---
    is very labourious to apply in loop calculations, 
    but is still useful in order to expose the fact that the trace
    of difference of two logarithms may actually depend on the common
    coefficient in their arguments, {\it i.e.}
\begin{equation}
\label{rhoinv}
    \Tr \bigl(\ln \chi^{2} r(M,\Lambda) - \ln \chi^{2} r(M_{0},\Lambda)\bigr) 
    \neq \Tr \bigl(\ln r(M,\Lambda) - \ln r(M_{0},\Lambda)\bigr) ,
\end{equation}
    where
$ M_{0} = M(0) $.
    The insertion of the coefficient
$ \chi $
    can be interpreted as an introduction of the integration measure in Eq.~\eqref{EAint}.
    Indeed, the change of the integration variable
\begin{equation*}
    b \to \chi b
\end{equation*}
    multiplies the propagator by $ \chi^2 $,
\begin{equation*}
    r(M,\Lambda) \to \chi^{2} r(M,\lambda)
\end{equation*}
    and the vertices
\begin{equation*}
    V_{k} \to \chi^{-k} V_{k} 
\end{equation*}
    by their corresponding powers of
$ \chi $.
    It is not hard to show that the contribution of the loop diagrams
    does not depend on 
$ \chi $,
    while the trace of the logarithm acquires a coefficient in its argument:
\begin{equation*}
    \Tr \bigl(\ln r(M,\Lambda) - \ln r(M_{0},\Lambda)\bigr) \to
    \Tr \bigl(\ln \chi^{2} r(M,\Lambda) - \ln \chi^{2} r(M_{0},\Lambda)\bigr) .
\end{equation*}
    The measure
$ \chi $, in its turn,
    can depend on
$ \Lambda $,
    and in this way 
    the choice of $ \chi $
    determines the renormalization scheme \cite{SF}.
    Moreover, as the functional integral is a product of integrals
    related to different parts of the spectrum of the quadratic form
    in the exponent,
    one can take
$ \chi $
    to be a product of different measures for each of the integrals.
    Or, in other words, a function of the quadratic form operator
($ M $ or
$ M_{0} $).

    These considerations show that the function in the argument of
    the logarithm, as a combination of 
$ r $
    and the measure
$ \chi $,
    can be varied to a significant extent, which enables us to arrange
    the overall expression in the logarithm trace to be well defined.
    More limitations on 
$ \chi $
    should be imposed in the process of
    renormalization, as will be illustrated further in the example of
    the Yang-Mills action.

\section{Heat kernel regularization}
    In order to render the expressions in the trace of 
    the logarithm and in the loop terms finite
    let us first restrict ourselves to 
    the class of Laplace transformations of the quadratic form operator
$ M $.
    We can write the regularized propagator and its logarithm
    as follows:
\begin{align}
\label{r1}
    r(M,\Lambda)
	= & \int_{0}^{\infty} \hat{r}(t,\Lambda) e^{-Mt} dt ,\\
\label{r2}
    l(M,\Lambda)
	= & \int_{0}^{\infty} \hat{l}(t,\Lambda) e^{-Mt} dt .
\end{align}
    The functions
$ r(M,\Lambda) $ and 
$ l(M,\Lambda) $
    must obey the conditions
\begin{align*}
    r(M,\Lambda) & \stackrel{\Lambda \to \infty}{\rightarrow} M^{-1} , \\
    l(M,\Lambda) & = \ln r(M,\Lambda) , \quad M \geq 0 .
\end{align*}
    Although the first argument here
    is an operator, most properties of
$ r $
    and
$ l $
    are fixed when it takes scalar (eigen) values.
    Thus, depending on the context, we will treat the argument
    in different senses.

    Besides, we require for 
$ r(M,\Lambda) $ and
$ l(M,\Lambda) $
    to be of
    a ``reasonable behaviour at zero'' in the coordinate representation.
    This implies a finite expression for the trace
\begin{equation}
\label{log1}
    \Tr \bigl(l(M,\Lambda) - l(M_{0},\Lambda) \bigr) = 
    \int \tr \int_{0}^{\infty} \hat{l}(t,\Lambda) (e^{-Mt}-e^{-M_{0}t})(x,y)
	 dt |_{x=y} d^{4}x 
\end{equation}
    and the divergence of the propagator
\begin{equation*}
    r(x,y) = \int_{0}^{\infty} \hat{r}(t,\Lambda) e^{-Mt}(x,y) dt 
\end{equation*}
    at least less than
$ (x-y)^{-2} $
    (in reality for finiteness of 8-like diagrams we also need to require 
    the existence of the limit of $ r(x,y) $ at equal arguments).

    Since the above divergences are related to the behaviour of
$ \hat{r}(t) $, $ \hat{l}(t) $
    in the vicinity of zero,
    we need to study the behaviour of the exponent
$ e^{-Mt} $
    near the origin.
    This exponent --- the heat kernel --- is defined by the equation
\begin{equation*}
    \frac{\partial e^{-Mt}}{\partial t} + M e^{-Mt} = 0 , \quad
	e^{-Mt} \stackrel{t\to 0}{\rightarrow} \delta^{mn}
	    \delta^{4}(x-y) 
\end{equation*}
    (here and further on $ m $ and $ n $ denote the indices of the operator $ M $ 
    related to the symmetries of the theory).
    We assume that the operator
$ M $
    obeys the limit
\begin{equation*}
    M_{0} = M|_{B=0} = -\partial_{\mu}\partial_{\mu} \delta^{mn} ,
\end{equation*}
    and that the heat kernel admits the following expansion near the origin
\begin{equation}
\label{hkexp}
    e^{-Mt} = e^{-M_{0}t}(a_{0} + a_{1}t + a_{2}t^{2} + \ldots) , \quad
	e^{-M_{0}t} = \frac{\delta^{mn}}{4\pi^{2}t^{2}}
	    e^{-\frac{(x-y)^{2}}{4t}} . 
\end{equation}
    Here the coefficients
$ a_{k} $
    must depend on
$ B $
    in such a way that
\begin{equation*}
    a_{0}|_{B=0} = \delta^{mn} , \quad a_{k}|_{B=0} = 0 , \quad k>0 
\end{equation*}
    (for a thorough discussion of the heat kernel please refer to the manual
\cite{HKman}).
    As an example, the Yang-Mills theory contains two quadratic forms
    with the following operators
\begin{gather*}
    M^{\text{YM}} = -\nabla\nabla \delta_{\mu\nu} - 2 F_{\mu\nu} , \quad
    M^{\text{ghost}} = - \nabla\nabla , \\
    \nabla_{\mu} = \partial_{\mu} + B_{\mu} , \quad
	F_{\mu\nu} = \nabla_{\mu}\nabla_{\nu} - \nabla_{\nu}\nabla_{\mu} ,
\end{gather*}
    and the coefficients
$ a_{k} $
    are defined by the equations
\begin{gather*}
    (x-y)^{\lambda} \nabla_{\lambda} a_{0} = 0 , \\
    k a_{k} + (x-y)^{\lambda} \nabla_{\lambda} a_{k} = -M a_{k-1} ,
\end{gather*}
    which yield
\begin{gather}
\label{acond1}
    a_{0}(x,x) = \delta^{mn} , \quad a_{1}(x,x)^{\{mn\}} =0 \\
\label{acond2}
    [a^{\text{YM}}_{2}(x,x)]^{mm}
	= - \frac{5}{12} \frac{C_{2}}{4\pi^{2}}	F_{\mu\nu}^{2} , \quad 
    [a^{\text{ghost}}_{2}(x,x)]^{mm} = \frac{1}{48} \frac{C_{2}}{4\pi^{2}}
	F_{\mu\nu}^{2} .
\end{gather}

    Taking into account conditions
\eqref{acond1} and
\eqref{acond2}
    one can conclude that the first coefficient that contributes to the 
    logarithm
\eqref{log1}
    with equal arguments is
$ a_{2} $:
\begin{multline}
\label{trlog}
    \Tr \bigl(l(M,\Lambda) - l(M_{0},\Lambda) \bigr) = \\
	= \int \int_{0}^{\infty} \hat{l}(t,\Lambda) 
    \frac{1}{4\pi^{2}t^{2}}e^{-\frac{(x-y)^{2}}{4t}} \bigl(
    (a_{0}+a_{1}t+a_{2}t^{2}+\ldots)^{mm} - \delta^{mm} \bigr)dt
	|_{x=y} d^{4}x =\\
	= \frac{1}{4\pi^{2}} \int [a_{2}(x,x)]^{mm}d^{4}x
	    \int_{0}^{\infty} \hat{l}(t,\Lambda) dt + \ldots
    = A_{2} l(M,\Lambda)|_{M=0} .
\end{multline}
    Here we have denoted
\begin{equation*}
    A_{k} = \frac{1}{4\pi^{2}} \int [a_{k}(x,x)]^{mm} d^{4}x\,, 
\end{equation*}
    and assumed that the integration over
$ t $
    and the limit
$ x=y $
    can be interchanged.
    Now let us take a look at the possible divergences of the integral
\begin{equation}
\label{lint}
    \int_{0}^{\infty} \hat{l}(t,\Lambda) dt = l(M,\Lambda)|_{M=0}.
\end{equation}
    It can diverge at the infinity of
$ t $
    if
$ l(M,\Lambda) $
    indefinitely grows at zero.
    This type of divergences can be eliminated by introducing an
    infrared parameter
$ \mu $
    (the renormalization point), {\it e.g.} via shifting
\begin{equation*}
    M \to M+\mu^{2}
\end{equation*}
    (if the theory is massive
$ \mu^{2} $
    can be extracted directly from $ M $  while keeping
    it positive).
    Then, from the properties of the Laplace transformation it follows that
\begin{equation*}
    \Tr \bigl(l(M+\mu^{2},\Lambda) - l(M_{0}+\mu^{2},\Lambda) \bigr)  
	= A_{2} \int_{0}^{\infty} l(t,\Lambda) e^{-\mu^{2} t} dt
	\simeq A_{2} l(\mu^{2},\Lambda) .
\end{equation*}

    The absence of divergence at zero in the integral
\eqref{lint}
    implies that function
\begin{equation*}
    l(M) = \int_{0}^{\infty} \hat{l}(t) e^{-Mt} dt
\end{equation*}
    is limited when
$ M \to \infty $.
    This statement holds for a class of preimage functions
$ \hat{l}(t) $
    which are ``regular'' at zero or are integrable by absolute value.
    It does not hold, for example, for generalized functions, although in this case,
    as we shall see later, the expansion
\eqref{hkexp}
    requires a special interpretation when calculating the trace.

    From the finite behaviour of 
$ l(M) $
    it follows that function
$ r(M) = \exp l(M) $
    does not tend to zero at infinity at all, 
    and in this way has a worse behaviour in the difference 
$ (x-y) $
    than
\begin{equation*}
    M_{0}^{-1} = \frac{\delta^{mn}}{4\pi^{2}(x-y)^{2}} .
\end{equation*}
    The boundary line is the function
\begin{equation*}
    \hat{l}_{\text{log}}(t) = \frac{1}{t} 
\end{equation*}
    --- in order for the trace of the logarithm to converge
$ \hat{l}(t) $
    should behave at zero better than
$ \hat{l}_{\text{log}}(t) $,
    although the corresponding propagator will now become more divergent.
    Vice versa, the Laplace preimages of
$ l=\ln r(M) $
    with
$ r(M)$  
    decreasing as
$ M^{-2} $
    and faster are given by derivatives of the delta-function,
    which behave at zero worse than    
$ \hat{l}_{\text{log}}(t) $.
    This is illustrated in 
\cite{HCD1},
\cite{HCD2},
    where the method of higher covariant derivatives
    is shown to work well with the loop terms, 
    but certain obstacles are found in the
    trace of the logarithm. 
    Outside the scope of the background
    field method a similar problem is discussed in
\cite{Slavnov77},
\cite{Martin},
\cite{BS}, also see the references therein.

    Let us take a look at what happens when
$ \hat{l}(t) $
    is a generalized function.
    For example, inverse Laplace transforms of functions
$ \ln \rho^{2} r(M,\Lambda) $ and
$ \ln r(M,\Lambda) $
    differ by
$ \delta(t)\ln\rho^{2} $,
    which allows us to write
\begin{multline*}
    \Tr\bigl(\ln \rho^{2} r(M,\Lambda) - \ln \rho^{2} r(M_{0},\Lambda)\bigr)
    - \Tr\bigl(\ln r(M,\Lambda) - \ln r(M_{0},\Lambda) \bigr) = \\
	=\ln \rho^{2} \int \tr \int_{0}^{\infty} \delta(t)
	    (e^{-Mt}-e^{-M_{0}t}) dt |_{x=y} d^{4}x .
\end{multline*}
    This gives us the trace of the difference of two identity operators
    which by common sense should vanish.
    But on the other hand, according to Eq.~\eqref{trlog} the expansion \eqref{hkexp} for
$ e^{-Mt} $
    yields,
\begin{multline*}
	\ln \rho^{2} \int \tr \int_{0}^{\infty} \delta(t)
	    (e^{-Mt}-e^{-M_{0}t}) dt |_{x=y} d^{4}x =\\
    =\ln \rho^{2} \int \tr \int_{0}^{\infty} \frac{\delta(t)}{4\pi^{2}t^{2}}
	e^{-\frac{(x-y)^{2}}{4t}} a_{2}(x,y) t^{2} dt |_{x=y} d^{4}x
    = A_{2} \ln \rho^{2} .
\end{multline*}    
    The expression in the outer integral 
\begin{equation*}
    \int_{0}^{\infty} \frac{\delta(t)}{4\pi^{2}}
	e^{-\frac{(x-y)^{2}}{4t}} a_{2}(x,y) dt
    = \begin{cases}
	\frac{1}{4\pi^{2}} a_{2}(x,x) , \quad & x=y ,\\
	0 , \quad &x \neq y
    \end{cases}
\end{equation*}
    is not continuous in
$ x,y $.
    As an operator kernel, it does not change the identity operator
$ e^{-Mt}|_{t=0} $,
    but at the same time it produces a nonzero trace.
    This fact may have a physical manifestation in terms of 
    breaking of the scale invariance of the logarithm
\eqref{rhoinv},
    however from the mathematical point of view it is just an incorrect
    interchange of the limit and the integration in Eq.~\eqref{trlog}.

    Concluding this section, we recap that a regularization such as
\eqref{r1},
\eqref{r2}
    with regular functions
$ \hat{r}(t)$, $\hat{l}(t) $
    is not suitable for the effective action in the background
    field method. 
    Meanwhile, admitting functions with a faster growth of the absolute value than
    that of
$ \ln M^{-1} $
    in Eq.~\eqref{r2} takes us out of the class of the Laplace transformations of
    regular preimages
$ \hat{l}(t) $,
    and thus
$ l(M,\Lambda) $
    becomes discontinuous in
$ x,y $
    and its trace not well defined.

    To finish this section
    we give two examples of functions
$ l(t) $
    and their corresponding Laplace preimages. 

\subsection{Example: cut-off in the Laplace transformation}
\label{exLapl}
    The first example is represented by a cut-off in the Laplace
    transformation at a position defined by a small parameter
$ 1/\Lambda^{2} $:
\begin{equation*}
    \hat{l}_{\text{cut}}(t,\Lambda) = \begin{cases}
		0, \quad t < 1/\Lambda^{2} , \\
		1/t, \quad 1/\Lambda^{2} \leq t .
	    \end{cases}
\end{equation*}
    This regularization taken in the
    above interpretation of the background field method was discussed in
\cite{LFC}.
    The regularized logarithm here looks as follows:
\begin{equation*}
    l(M,\Lambda) =\int_{0}^{\infty} \hat{l}_{\text{cut}}(t) e^{-Mt} dt
	= \int_{1/\Lambda^{2}}^{\infty} \frac{e^{-Mt}}{t} dt
	= \Ee_{1}(M/\Lambda^{2}) ,
\end{equation*}
    while for its trace the relation
\eqref{trlog}
    yields 
\begin{equation*}
    \Tr \bigl(l(M+\mu^{2},\Lambda) - l(M_{0}+\mu^{2},\Lambda) \bigr)  
	= A_{2} \Ee_{1}(\mu^{2}/\Lambda^{2}) .
\end{equation*}
    At small arguments the integral exponent
$ \Ee_{1} $ behaves as
\begin{equation*}
    \Ee_{1}(M/\Lambda^{2}) \simeq -\ln \frac{M}{\Lambda^{2}} - \gamma
	+ o(1) ,
\end{equation*}
    and so we get an infinite growth in the trace.
    This divergence, however, arises mainly due to the fact that at
$ \Lambda\to\infty $
\begin{equation}
\label{cut-off}
    l(M,\Lambda) \simeq -\ln \frac{M}{\Lambda^{2}} ,
\end{equation}
    that is, we are taking 
$ \Lambda^{2} M^{-1} $
    instead of
$ M^{-1} $
    for the propagator.

    On the other hand, when
$ M $
    goes to infinity we get an expansion
\begin{equation*}
    \Ee_{1}(M/\Lambda^{2}) \simeq e^{-M/\Lambda^{2}}
	\bigl(\frac{\Lambda^{2}}{M} + o(1)\bigr)
	    \stackrel{M\to\infty}{\rightarrow} 0 ,
\end{equation*}
    which prevents us from using the function
\begin{equation*}
    r(M,\Lambda) = \exp \{\Ee_{1}(M/\Lambda^{2})\}
	\stackrel{M\to\infty}{\simeq} I + o(1)
\end{equation*}
    as a regularized propagator in loop calculations.

\subsection{Example: Pauli-Villars regularization}
\label{exPV}
    The second example is the Pauli-Villars regularization 
\cite{PV}.
    In a simplified description it is a given by the Laplace transform of the
    function
\begin{equation*}
    \hat{l}_{\text{PV}}(t,\Lambda) = \frac{1-e^{-\Lambda^{2}t}}{t} ,
\end{equation*}
    which looks as follows
\begin{equation}
\label{PVtr}
    l(M,\Lambda) = \int_{0}^{\infty} \frac{1-e^{-\Lambda^{2}t}}{t}
	e^{-Mt} dt = \ln \frac{M+\Lambda^{2}}{M} .
\end{equation}
    An actual Pauli-Villars regularization includes several exponents
    with different weights, but its resulting behaviour at infinities in
$ M $ and
$ \Lambda $
    is the same as in the above example.

    The corresponding trace of the logarithm is expressed as an
    elementary function:
\begin{equation*}
    \Tr \bigl(l(M+\mu^{2},\Lambda) - l(M_{0}+\mu^{2},\Lambda) \bigr)  
	= A_{2} \ln \frac{\Lambda^{2}}{\mu^{2}} .
\end{equation*}
    Although it grows at
$ \Lambda \to \infty $,
    however, again, this growth is related to the growth of the multiplier
    at the propagator in the argument of the logarithm:
\begin{equation}
\label{PVas}
    l(M,\Lambda) \stackrel{\Lambda\to\infty}{\simeq}
	\ln \frac{\Lambda^{2}}{M} .
\end{equation}
    At the same time when 
$ M $
    goes to infinity we have
\begin{equation*}
    r(M,\Lambda) = \exp l(M,\Lambda)
	= \exp \{\ln\frac{M+\Lambda^{2}}{M}\} \simeq I + o(1) ,
\end{equation*}
    which means that the remark in the previous example
    about the bad behavior of the propagator at large
$ M $
    is also applicable here.

\section{Restriction of the integration domain}
    An alternative approach to regularize the integral in Eq.~\eqref{EAint}
    can be a (formal) restriction of the domain of functions over
    which the integration is performed.
    Let us take into account only those functions which
    obey an inequality
\begin{equation}
\label{bmb}
    \int (b,Mb) d^{4}x \leq \Lambda^{2} \int (b,b) d^{4}x
\end{equation}
    and its consequence
\begin{equation*}
    \int (b, (\ln M) b)  d^{4}x \leq \ln \Lambda^{2} \int (b,b) d^{4}x ,
\end{equation*}
    the latter being valid if $ M $ is positive.
    Then the regularized propagator and its logarithm can be
    represented as the expressions
\begin{align*}
    r(M,\Lambda) =& \begin{cases}
	M^{-1}, \quad &|M| \leq \Lambda^{2} ,\\
	0 , \quad &\Lambda^{2} < |M| ,
		    \end{cases} \\
    l(M,\Lambda) = & \begin{cases}
	-\ln |M| , \quad &|M| \leq \Lambda^{2} ,\\
	0 , \quad &\Lambda^{2} <|M| .
		    \end{cases} 
\end{align*}
    Indeed, let
$ P^{\Lambda} $
    be a projector to the spectral subspace of the operator
$ M $ which corresponds to the part of the spectrum between 0 and
$ \Lambda^{2} $.
    Then the integral over functions satisfying
\eqref{bmb}
    can be written in terms of this projector and transformed as follows:
\begin{align}
\nonumber
    \int W(b) & \exp \{\frac{i}{2}bMb\} \prod_{P^{\Lambda}b=b} \chi \delta b 
    = \int W(P^{\Lambda}b) \exp \{\frac{i}{2}bP^{\Lambda}MP^{\Lambda}b\}
	\prod_{P^{\Lambda}b=b} \delta \chi b = \\
\nonumber
    &= W(\frac{1}{i\chi}\frac{\delta}{\delta j}) \int
	\exp \{\frac{i}{2}\tilde{b} P^{\Lambda}
	    \frac{M}{\chi^{2}} P^{\Lambda}\tilde{b} +i\tilde{b}P^{\Lambda}j\} 
	\prod_{P^{\Lambda}\tilde{b}=\tilde{b}} \delta \tilde{b} = \\
\nonumber
    &= (\Det \chi^{-2}P^{\Lambda}M)^{-1/2}
	    W(\frac{1}{i\chi}\frac{\delta}{\delta j}) 
	\exp \{-\frac{i}{2}jP^{\Lambda}\chi^{2} M^{-1}P^{\Lambda}j\}|_{j=0} =\\
\label{funint}
    &= \exp\{\frac{1}{2}\Tr\ln \chi^{2} r(M,\Lambda)\}
	W(\frac{\delta}{i\delta j}) \exp \{-\frac{i}{2}jr(M,\Lambda)j\}|_{j=0} .
\end{align}
    Similarly to the case with the functional
    integral over the full space of functions
    this relation is proved for polynomial forms
$ W(b) $ 
    (see the definition of the functional integral in
\cite{SF}).
    The determinant
$ \Det P^{\Lambda}M $
    is understood as a product of eigenvalues with an account for
    multiplicities over the part of the spectrum between 0 and
$ \Lambda^{2} $.
    Besides, we have introduced the scalar measure
$ \chi $
    which as we mentioned above only contributes to the trace of 
    logarithm, but not to loop calculations.

    The functions
$ r(M,\Lambda) $ and
$ l(M,\Lambda) $
    are not continuous in
$ M $, and therefore they are not in the class of Laplace
    transformations. Instead, they can be represented as
    Fourier images:
\begin{align*}
    r(M,\Lambda) =& \frac{i}{\pi} \int \Si(\Lambda^{2}t) e^{-iMt} dt , \\
    \ln \chi^{2} r(M,\Lambda) = l(\chi^{-2}M,\Lambda)
	=& \frac{1}{\pi} \int \bigl( \frac{\Si(\Lambda^{2}t)}{t}
	- \frac{\sin\Lambda^{2}t}{t}\ln \frac{\Lambda^{2}}{\chi^{2}} \bigr)
	    e^{-iMt} dt , \\
    P^{\Lambda}(M) =& \frac{1}{\pi} \int \frac{\sin\Lambda^{2}t}{t}
	e^{-iMt} dt ,
\end{align*}
    where the exponent
$ e^{-iMt} $
    is defined by the equation
\begin{equation*}
    \frac{\partial e^{-iMt}}{i\partial t} + M e^{-iMt} = 0 ,
	\quad e^{-iMt} \stackrel{t\to\pm 0}{\longrightarrow}
	    \delta^{mn} \delta^{4}(x-y).
\end{equation*}
    This type of exponent can be derived from the expansion
\eqref{hkexp}
    by substituting
$ t \to it $:
\begin{equation}
    e^{-iMt} = e^{-iM_{0}t}(a_{0} + i a_{1}t - a_{2}t^{2} + \ldots) , \quad
	e^{-iM_{0}t} = \frac{-\delta^{mn}}{4\pi^{2}t^{2}}
	    e^{i\frac{(x-y)^{2}}{4t}} .
\end{equation}
    Let us mention that as the function
$ r(M,\Lambda) $
    vanishes everywhere starting from the point
$ \Lambda^{2} $,
    the corresponding operator in the coordinate
    representation is regular at equal arguments:
\begin{equation*}
    r(x,y) \simeq \frac{J_{0}(\Lambda|x-y|)-1}{4\pi^{2}(x-y)^{2}} a_{0}(x,y)
	+ o(1) \simeq \frac{\Lambda^{2}}{4\pi^{2}} \delta^{mn} + o(1) .
\end{equation*}
    The trace of the logarithm
$ l(M,\Lambda) $
    (for the Yang-Mills field)
    is calculated via an equation similar to
Eq.~\eqref{trlog},
    which is based on the cancellation of the power
$ t^{2} $
    in front of the coefficient
$ a_{2} $
    with that of the denominator of the kernel
$ e^{iM_{0}t} $.
    After the introduction of an infrared parameter
$ \mu $
    we get
\begin{multline}
\label{trlogc}
    \Tr \bigl(\ln \chi^{2} r(M+\mu^{2},\Lambda)
	- \ln \chi^{2} r(M_{0}+\mu^{2},\Lambda) \bigr) =\\
	= \frac{1}{\pi} \int \tr \int \bigl( \frac{\Si(\Lambda^{2}t)}{t}
    - \frac{\sin\Lambda^{2}t}{t}\ln \frac{\Lambda^{2}}{\chi^{2}} \bigr)
	(e^{-i(M+\mu^{2})t} - e^{-i(M_{0}+\mu^{2})t})dt|_{x=y} d^{4}x = \\
    = \frac{1}{4\pi^{2}} \int
	    \bigl(Q_{2}(x-y) - \ln\frac{\Lambda^{2}}{\chi^{2}}q_{2}(x-y)\bigr)
	[a_{2}(x,y)]^{mm} |_{x=y} d^{4}x
    = A_{2} l(\frac{\mu^{2}}{\chi^{2}},\Lambda) =\\
	= A_{2} \ln \frac{\chi^{2}}{\mu^{2}} .
\end{multline}
    The explicit form of the functions
$ q_{2}(x) $ and
$ Q_{2}(x) $
    is not relevant, the answer obtained by the property of
    Fourier transform.
    We still provide the expressions for these functions in order to stress
    that the change of the order of integration over
$ t $
    and the limit 
$ x=y $
    is a correct operation:
\begin{equation*}
    q_{2}(x) = J_{0}(\sqrt{(\Lambda^{2}-\mu^{2})x^{2}}) , \quad
    Q_{2}(x) = \int_{\mu^{2}}^{\Lambda^{2}}
	J_{0}(\sqrt{(k-\mu^{2})x^{2}}) \frac{dk}{k} .
\end{equation*}
    Despite a manifest coefficient of
$ \ln \Lambda^{2} $
    in
\eqref{trlogc},
    at $ x=0 $
    this logarithm is cancelled,
    and one finds
\begin{equation*}
    Q_{2}(0) - \ln \frac{\Lambda^{2}}{\chi^{2}} \, q_{2}(0)
	= \ln \frac{\Lambda^{2}}{\mu^{2}}
	    - \ln\frac{\Lambda^{2}}{\chi^{2}} = \ln \frac{\chi^{2}}{\mu^{2}} .
\end{equation*}

    The expression
\eqref{trlogc}
    shows that within the current method of
    calculation the trace of the logarithm 
    does not directly depend on the regularization parameter
$ \Lambda $ (more precisely, it does not grow with
$ \Lambda $).
    From that expression it is also evident that a multiplication of the argument of the logarithm
    by the constant
$ \chi^{2} $ mentioned in
    Eq.~\eqref{rhoinv}
    adds one more term to the trace:
\begin{equation*}
    \Tr \bigl(\ln \chi^{2} r(M) - \ln \chi^{2} r(M_{0}) \bigr) =
    \Tr \bigl(\ln r(M) - \ln r(M_{0}) \bigr) + A_{2} \ln \rho^{2} .
\end{equation*}
    This behaviour of the trace of the logarithm is related to the fact
    that upon extracting the terms with a coefficient of
$ \ln\chi^{2} $,
    instead of cancelling the traces of the identity operators,
    we should rather cancel the traces of the projectors which count the
    difference of the respective ``numbers of eigenfunctions'' of the operators
$ M $ and
$ M_{0} $.
    It is also natural that this difference does not vanish as
$ \Lambda \to \infty $,
    even though
$ M $ and
$ M_{0} $
    both operate in the ``same space''.

    The expression
\eqref{trlogc} also reveals
    that both the effective action and the renormalization process depend on the initial
    choice of the integration measure
$ \chi $.
    The latter should be chosen in such a way as to compensate
    for the contributions growing with
$ \Lambda $
    in the loop terms by an addition to the trace, 
    and this way to derive a finite expression
    for the renormalized effective action.
    One particular condition of such a compensation is the equality
\begin{multline}
\label{rcond}
    \frac{\delta}{\delta B} \bigl(
	\ln\int\exp \{\frac{i}{2}b \frac{M}{\chi^{2}} b\}
	    \prod_{P^{\Lambda}b=b}\delta b
	-\ln\int\exp \{\frac{i}{2}b \frac{M_{0}}{\chi^{2}} b\}
	    \prod_{P_{0}^{\Lambda}b=b}\delta b
	\bigr) \simeq \\
    \simeq \frac{i}{2\chi^{2}} \int b\frac{\delta M^{\Lambda}}{\delta B}b
	\exp \{\frac{i}{2}b\frac{M}{\chi^{2}}b\} \prod_{P^{\Lambda}b=b}\delta b
    \cdot \bigl(\int\exp\{\frac{i}{2}b\frac{M}{\chi^{2}}b\}
	\prod_{P^{\Lambda}b=b}\delta b \bigr)^{-1}
\end{multline}
    which interrelates the primary divergences in diagrams with
    different numbers of loops (an analogue of the Ward identity).
    The usual rule for variation of the logarithm is not applicable here.
    The reason is that it is not the argument of the logarithm that is varied, but rather
    the spectrum multiplicity, which is a \emph{coefficient} at the logarithm, is.
    Thus the LHS above is equal (up to an infrared shift by
$ \mu^{2} $)
    to a variation of the trace \eqref{trlogc} with respect to the background field,
\begin{equation*}
    \text{LHS} \simeq \ln \frac{\chi^{2}}{\mu^{2}} \,
	\frac{\delta A_{2}}{\delta B} .
\end{equation*}
    While the RHS, although it does not depend on
$ \chi $,
    grows with
$ \Lambda $ as
\begin{equation*}
    -\frac{1}{2} \Tr \frac{\delta M}{\delta B} r(M,\Lambda) \simeq
    -\frac{1}{2} \Tr \frac{\delta M}{\delta B} a_{1} Q_{2}(0)
	\simeq - \frac{1}{2} \ln \frac{\Lambda^{2}}{\mu^{2}} 
	    \Tr \frac{\delta M}{\delta B} a_{1} .
\end{equation*}
    To evaluate the RHS we can use the following expansion of the propagator
$ r(M,\Lambda) $
    in powers of
$ (x-y) $:
\begin{align*}
    r(M & + \mu^{2}, \Lambda)
	= \frac{i}{\pi} \int \Si(\Lambda^{2}t) e^{-iMt-i\mu^{2}t} dt = \\
	=& \frac{-i}{4\pi^{2}\pi} \int \Si(\Lambda^{2}t)
	    e^{i\frac{(x-y)^{2}}{4t}-i\mu^{2}t}
		(a_{0}+ia_{1}t-a_{2}t^{2} +\ldots) \frac{dt}{t^{2}} = \\
	=& \frac{1}{4\pi^{2}} (Q_{1}(x-y)a_{0}(x,y) + Q_{2}(x-y) a_{1}(x,y)
	    + Q_{3}(x-y) a_{2}(x,y) + \ldots ),
\end{align*}
    where
\begin{align*}
    Q_{1}(x) = &2\int_{\mu^{2}}^{\Lambda^{2}} \sqrt{\frac{k-\mu^{2}}{x^{2}}}
	J_{1}(\sqrt{(k-\mu^{2})x^{2}}) \frac{dk}{k}
	    \simeq \Lambda^{2}-\mu^{2} - \mu^{2}\ln\frac{\Lambda^{2}}{\mu^{2}}
		+ o(1) ,\\
    Q_{3}(x) = &\frac{1}{2} \int_{\mu^{2}}^{\Lambda^{2}}
	\sqrt{\frac{x^{2}}{k-\mu^{2}}}
	J_{1}(\sqrt{(k-\mu^{2})x^{2}}) \frac{dk}{k}
	    \simeq \frac{1}{4} x^{2} \ln \frac{\Lambda^{2}}{\mu^{2}} + o(x^{2}),
\end{align*}
    and
$ Q_{2}(x) $
    is as before.

    In the Yang-Mills theory example both of the quadratic form operators
    obey the relation\footnote{
Although it looks quite natural, the author is only aware of a ``straightforward'' proof
of this relation, which takes half a page of 
$ \nabla $-algebra transformations per each operator.
}
\begin{equation}
\label{YMcond}
    \Tr \frac{\delta M}{\delta B} a_{1} = - \Tr \frac{\delta a_{2}}{\delta B} ,
\end{equation}
    and Eq.~\eqref{rcond} gives
$ \chi = \Lambda $.
    Thus, the logarithm trace together with the integration measure
    yield the well-known leading divergent term in the effective action
\begin{align*}
    \EA(B) = & \frac{1}{g^{2}} S_{\text{cl}}
	+ \frac{1}{2} \ln \frac{\Lambda^{2}}{\mu^{2}} A^{\text{YM}}_{2}
	- \ln \frac{\Lambda^{2}}{\mu^{2}} A_{2}^{\text{ghost}} + \ldots =\\
    &= \frac{1}{g^{2}} S_{\text{cl}}
	- \frac{11}{48} \frac{C_{2}}{4\pi^{2}} \ln \frac{\Lambda^{2}}{\mu^{2}}
	  S_{\text{cl}} + \ldots .
\end{align*}
    Being rather difficult to apply even at the two-loop approximation, the
    scheme described in this section together with Eq.~\eqref{rcond} 
    provide an important hint for a possible application of the
    integration measure.

\section{Heat kernel. Extended version}

    Let us consider a function
$ \Omega_{M}(\lambda) $
    --- the density of the number of eigenvectors at a spectral point
$ \lambda $.
    Or, in the other words, the (somewhat re-scaled) number of eigenvectors
    with the eigenvalues sitting in a spectral interval around the point
$ \lambda $,
    divided by the length of the interval.
    For example, for the operator
$ M_{0} = - \partial^{2} $
    in a 4-dimensional space the number of eigenvectors in the interval
$ [\lambda , \lambda+d\lambda] $
    is proportional to
$ \lambda d\lambda $,
    and so we can write
\begin{equation*}
    \Omega_{M_{0}}(\lambda) = c\lambda ,
\end{equation*}
    where
$ c $
    is a coefficient of dimension
$ \lambda^{-2} $
    (it has to be of this dimension since the \emph{number} of eigenvectors
$ \Omega_{M_{0}}(\lambda) \, d\lambda $
    is dimensionless).
    Further, we assume that the spectrum of operator
$ M $
    has the same behaviour at infinity as the spectrum of
$ M_{0} $ does,
    and in this way we can introduce a difference function
$ \omega $
    vanishing at infinity,
\begin{equation*}
    \Omega_{M}(\lambda) = c\lambda + \omega(B,\lambda) .
\end{equation*}
    This function allows us to write formal expressions for the re-scaled
    difference of the numbers of eigenvalues of
$ M $ and
$ M_{0} $
    in the interval
$ [\lambda',\lambda''] $
\begin{equation*}
    \int_{\lambda'}^{\lambda''} \omega(\lambda) \, d\lambda
\end{equation*}
    and then for the traces of operators
$ l(M) $ and
$ l(M_{0}) $
    (valid for some set of functions $ l $):
\begin{equation}
\label{lom}
    \Tr l(M) - \Tr l(M_{0}) = \int l(\lambda) \omega(\lambda) \, d\lambda .
\end{equation}
    Although we know little about the density
$ \omega(\lambda) $
    (direct comparison of
Eq.~\eqref{lom}
    with the extention of
Eq.~\eqref{trlog}
    reveals it as a sum of derivatives of
$ \delta $-functions with coefficients
$ A_{k} $,
    which seems to be incorrect), our main purpose
    is the expression for the contribution of the measure
$ \chi $
    to the effective action
\eqref{EAint}. 
    For variables obeying the Bose-Einstein statistics
    this expression looks as follows,
\begin{align}
\label{chicontr}
    \EA(B) = & \ln \int\exp\{iS(B,b)\} \prod\chi\delta b
		-\ln \int\exp\{iS(0,b)\} \prod\chi\delta b = \\
\nonumber
    = & \ln \int\exp\{iS(B,b)\} \prod \delta b
		-\ln \int\exp\{iS(0,b)\} \prod \delta b + \\
\nonumber
	&+ \int \omega(\lambda) \ln \chi(\lambda) \, d\lambda .
\end{align}
    Here we also assume that
$ \chi $
    can be different for those components of the variation
$ \delta b $
    which correspond to different parts of the spectrum of quadratic
    form in the functional integral.

    The contribution to the effective action of the integration measure
$ \chi $ together with the expression
\eqref{chicontr}
    suggest us how to overcome the difficulties with the heat kernel
    regularization described in Section 1.
    The divergence in the trace of the logarithm in Sections
\ref{exLapl} and
\ref{exPV}
    does not stem from an inefficient decrease of the expressions
    in the integrals of the type
\eqref{lom},
    but rather from the multiplication of the propagator,
    of which we are taking the
    logarithm, by the regularization parameter
$ \Lambda $
\eqref{cut-off},
\eqref{PVas}.
    At the same time, the functional integral in the effective action is itself
    defined up to the measure
$ \chi $,
    which only enters the trace terms.
    Hence it is our choice to change the quadratic form as
$ M \to r^{-1}(M,\Lambda) $
    in such a way as to render the loop terms finite and 
    to compensate for the divergent trace of the logarithm
    by measure terms as those in Eqs.
\eqref{trlogc} and
\eqref{chicontr}.

    More precisely, the method of higher covariant derivatives
\cite{HCD1},
\cite{HCD2}
    multiplies
$ M $
    by a polynomial of degree
$ n $,
\begin{equation*}
    M \to r^{-1}(M,\Lambda) = M p(\frac{M}{\Lambda^{2}})
\end{equation*}
    with a fixed behaviour at infinity and at zero:
\begin{align*}
    p(\tau) &\simeq \tau^{n} , \quad \tau \to \infty, \\
    p(\tau) &= 1 , \quad \tau = 0, 
\end{align*}
    which makes the loop terms finite.
    At the same time, the reverse Laplace transformation of
$ l(M) = -\ln M p(\frac{M}{\Lambda^{2}}) $
    behaves at zero as
\begin{equation*}
    \hat{l}(t) \simeq \frac{1+n}{t} , \quad t \to 0
\end{equation*}
    leading to a divergent integral over
$ t $ in Eq.\,\eqref{trlog}.
    At this point we can take
$ \chi $
    to be a function of
$ \lambda $
(but with constant asymptotics at
$ \Lambda \to\infty $):
\begin{equation*}
    \chi^{2}(\lambda) = (\lambda+\mu^{2}+\Lambda^{2})
	p(\frac{\lambda+\mu^{2}}{\Lambda^{2}})
	\stackrel{\Lambda\to\infty}{\simeq} \Lambda^{2} + O(\Lambda^{-1})
\end{equation*}
    and find the following contributions of the logarithm's trace and the measure
\begin{multline}
\label{mln}
    -\frac{1}{2} \int_{0}^{\infty} \ln (\lambda+\mu^{2})
	    f(\frac{\lambda+\mu^{2}}{\Lambda^{2}}) \:
	\omega(\lambda) d\lambda
    + \int_{0}^{\infty} \ln \chi(\lambda)
	\omega(\lambda) d\lambda = \\
    = -\frac{1}{2} \int_{0}^{\infty}
	    \ln\frac{(\lambda +\mu^{2})
	    f(\frac{\lambda+\mu^{2}}{\Lambda^{2}})}{\chi^{2}(\lambda)}
	\: \omega(\lambda) d\lambda 
    = -\frac{1}{2} \int_{0}^{\infty} \ln
	\frac{\lambda+\mu^{2}}{\lambda+\mu^{2}+\Lambda^{2}}
	\: \omega(\lambda) d\lambda = \\
    = - \frac{1}{2} \Tr \ln \frac{M+\mu^{2}}{M+\mu^{2}+\Lambda^{2}} .
\end{multline}
    This expression (taking the Bose-Einstein power
    coefficient $-1/2$ and the infrared term
$ \mu^{2} $ into account) coincides 
    with the expression \eqref{PVtr} calculated with the Pauli-Villars method.

    Now, to conclude, let us write out the main properties of the propagator.
    First of all, the latter is a Laplace transform,
\begin{multline*}
    \frac{1}{(M+\mu^{2}) p(\frac{M}{\Lambda^{2}})}
	= \int_{0}^{\infty} r(t) e^{-Mt} dt = \\
	= \frac{1}{4\pi^{2}} (L_{1}(x-y)a_{0}(x,y)
	    + L_{2}(x-y) a_{1}(x,y)
	    + L_{3}(x-y) a_{2}(x,y) + \ldots ) .
\end{multline*}
    Then, assuming that the roots
$ \tau_{k} $ of
$ p(\tau) $
    do not coincide, it can be transformed as follows,
\begin{align*}
    \frac{1}{M p(\frac{M}{\Lambda^{2}})} 
	=& \frac{\Lambda^{2n}\tau_{1}\ldots \tau_{n}}{M(M+\tau_{1}\Lambda^{2})
	    \ldots (M+\tau_{n}\Lambda^{2})} = \\
    &= \frac{1}{M} - \frac{d_{1}}{M+\tau_{1}\Lambda^{2}} - \ldots
	- \frac{d_{n}}{M+\tau_{n}\Lambda^{2}} ,
\end{align*}
    where
\begin{equation*}
    d_{k} = \frac{\tau_{1}\ldots\tau_{k-1}\tau_{k+1}\ldots\tau_{n}}{
	(\tau_{k}-\tau_{1})\ldots(\tau_{k}-\tau_{k-1})(\tau_{k}-\tau_{k+1})
	    \ldots (\tau_{k}-\tau_{n})}\,,
\end{equation*}
    and, in particular,
\begin{equation*}
    \sum_{k} d_{k} =1 ,\quad \sum_{k}\tau_{k}d_{k} = 0 , \quad
	\sum_{k}\tau_{k}^{-1} = \sum_{k} \tau_{k}^{-1} d_{k} .
\end{equation*}
    This allows us to write the first terms of the expansion of
$ L_{1,2,3} $
    around zero:
\begin{align*}
    L_{1} = & \int_{0}^{\infty} e^{-\frac{x^{2}}{4t}}
	(e^{-\mu^{2}t} -\sum_{k} d_{k} e^{-\Lambda_{k}^{2}t})\frac{dt}{t^{2}}
	= \frac{4}{x} \bigl(\mu \KK_{1}(\mu x) - \sum_{k} d_{k} \Lambda_{k}
	    \KK_{1}(\Lambda_{k} x)\bigr) = \\
	& = \frac{4}{x^{2}} (1-\sum_{k} d_{k}) + \mu^{2}\ln \mu^{2} x^{2}
	    - \sum_{k} d_{k}\Lambda_{k}^{2} \ln\Lambda_{k}^{2} x^{2}+ o(1)=\\
	&= \mu^{2}\ln\mu^{2} - \sum_{k}d_{k}\Lambda_{k}^{2}\ln\Lambda_{k}^{2}
	    + o(1) \stackrel{\Lambda\to\infty}{\simeq} 
	    -\mu^{2}\ln\frac{\Lambda^{2}}{\mu^{2}} ,
\end{align*}
    where
\begin{equation*}
    \Lambda_{k}^{2} = \mu^{2} + \tau_{k}\Lambda^{2} , \quad
	\sum_{k}d_{k}\Lambda_{k}^{2} = \mu^{2} .
\end{equation*}
    Not only does the coefficient at
$ x^{-2} $ vanish, but so does the coefficient at 
$ \ln x $,
    which ensures that 8-like diagrams are defined correctly.
    Then we can write
\begin{align*}
    L_{2} = & \int_{0}^{\infty} e^{-\frac{x^{2}}{4t}}
	(e^{-\mu^{2}t} -\sum_{k} d_{k} e^{-\Lambda_{k}^{2}t})\frac{dt}{t}
	= 2\bigl(\KK_{0}(\mu x) - \sum_{k} d_{k} \KK_{0}(\Lambda_{k} x)\bigr)
	    = \\
	& = -\ln \mu^{2}x^{2} + \sum_{k} d_{k}\ln\Lambda_{k}^{2} x^{2}+o(1)
	    =\ln \frac{\Lambda^{2}}{\mu^{2}} + o(1) ,
\end{align*}
    which allows us to compare the coefficient at
$ a_{1} $
    with the divergence in the RHS of Eq.\,\eqref{mln}
    and this way to check the renormalization condition
\eqref{rcond}. 
    Finally,
\begin{align*}
    L_{3} = & \int_{0}^{\infty} e^{-\frac{x^{2}}{4t}}
	(e^{-\mu^{2}t} -\sum_{k} d_{k} e^{-\Lambda_{k}^{2}t})dt = \\
	& = x \mu^{-1}\KK_{1}(\mu x)
	    - x \sum_{k}d_{k}\Lambda_{k}^{-1} \KK_{1}(\Lambda_{k}x) =\\
    &= \mu^{-2} -\sum_{k}d_{k}(\mu^{2}+\tau_{k} \Lambda^{2})^{-1} 
	+\frac{x^{2}}{4}(\ln\mu^{2}-\sum_{k}d_{k}\ln\Lambda_{k}^{2})+o(x^{2}).
\end{align*}
    Specific conditions can be imposed on the roots of
$ p(\tau) $
    in the process of calculation of two- and higher-loop terms, but this
    is a subject of a more thorough investigation.

\section*{Acknowledgments}
    The author is grateful to S.~Derkachov, A.~Pronko, P.~A.~Bolokhov
    and L.~D.~Faddeev for discussions. The work is partially supported by
RFBR grants 11-01-00570, 12-01-00207
    and the programme ``Mathematical problems of nonlinear dynamics'' of RAS.


\begin{thebibliography}{00}
\bibitem{DeWitt}
B.~S.~DeWitt,
``Quantum Theory of Gravity. 2. The Manifestly Covariant Theory, 3.
Applications of the Covariant Theory.''
  Phys.\ Rev.\  {\bf 162} (1967) 1195, 1239.

\bibitem{Abbott}
L.~F.~Abbott,
  ``The Background Field Method Beyond One Loop,''
  Nucl.\ Phys.\ B {\bf 185} (1981) 189, and references therein.

\bibitem{FSep}
L.~D.~Faddeev,
  ``Separation of scattering and selfaction revisited,''
    arXiv:1003.4854 [hep-th].

\bibitem{LFC}
    L.~D.~Faddeev,
``Mass in Quantum Yang-Mills Theory: Comment on a Clay Millenium problem,''
 arXiv:0911.1013 [math-ph].

\bibitem{JO}
    I.~Jack and H.~Osborn,
  ``Two Loop Background Field Calculations For Arbitrary Background Fields,''
  Nucl.\ Phys.\ B {\bf 207} (1982) 474.

\bibitem{HCD1}
    A.~A.~Slavnov,
    ``Invariant regularization of gauge theories,''
    Teor.\ Mat.\ Fiz.\  {\bf 13} (1972) 174.

\bibitem{HCD2}
    B.~W.~Lee and J.~Zinn-Justin,
  ``Spontaneously broken gauge symmetries ii. perturbation theory and
  renormalization,''
    Phys.\ Rev.\ D {\bf 5} (1972) 3137
    [Erratum-ibid.\ D {\bf 8} (1973) 4654].

\bibitem{SF}
  L.~D.~Faddeev and A.~A.~Slavnov,
``Gauge Fields. Introduction To Quantum Theory,''
Front.\ Phys.\  {\bf 50} (1980) 1, [Front.\ Phys.\  {\bf 83} (1990) 1].

\bibitem{HKman}
    D.~V.~Vassilevich,
  ``Heat kernel expansion: User's manual,''
  Phys.\ Rept.\  {\bf 388} (2003) 279
  [hep-th/0306138].

\bibitem{Slavnov77}
  A.~A.~Slavnov,
``The Pauli-Villars Regularization for Nonabelian Gauge Theories,''
  Teor.\ Mat.\ Fiz.\  {\bf 33} (1977) 210.

\bibitem{Martin}
C.~P.~Martin and F.~Ruiz Ruiz,
  ``Higher covariant derivative Pauli-Villars regularization does not lead to
  a consistent QCD,''
    Nucl.\ Phys.\ B {\bf 436} (1995) 545
      [hep-th/9410223].

\bibitem{BS}
  T.~D.~Bakeyev and A.~A.~Slavnov,
  ``Higher covariant derivative regularization revisited,''
    Mod.\ Phys.\ Lett.\ A {\bf 11} (1996) 1539
    [hep-th/9601092].

\bibitem{PV}
    W.~Pauli and F.~Villars, ``On the Invariant Regularization in Relativistic
    Quantum Theory,''
    Rev. Mod. Phys {\bf 21} (1949) 434-444.


\end{thebibliography}
\end{document}